\documentclass[preprint,prb,amsmath,amssymb,amsfonts,superscriptaddress,floatfix,showpacs]{revtex4}
\usepackage{graphicx}% Include figure files
\usepackage[dvips]{color}
\usepackage{dcolumn}% Align table columns on decimal point
\usepackage{bm}% bold math
\usepackage{rotating} % rotate figure
\usepackage{color}

\begin{document}
\title{Rare Earth Monopnictides and Monochalcogenides from First Principles: Towards an Electronic Phase Diagram of Strongly Correlated Materials}

\author{L. Petit}
% \email{lpetit@phys.au.dk}
 \email{leon.petit@stfc.ac.uk}
 \affiliation{Department of Physics and Astronomy, Aarhus University, DK-8000 Aarhus C, Denmark}
 \affiliation{ Daresbury Laboratory, Daresbury, Warrington WA4 4AD, UK }
\author{R. Tyer}
 \affiliation{ Daresbury Laboratory, Daresbury, Warrington WA4 4AD, UK }
\author{Z. Szotek}
 \affiliation{ Daresbury Laboratory, Daresbury, Warrington WA4 4AD, UK }
\author{W. M. Temmerman}
 \affiliation{ Daresbury Laboratory, Daresbury, Warrington WA4 4AD, UK }
\author{A. Svane}
 \affiliation{Department of Physics and Astronomy, Aarhus University, DK-8000 Aarhus C, Denmark}

%\date{\today}

\begin{abstract} 

{
We present results of an ab-initio study of the electronic structure of 140 rare earth compounds. Specifically we 
predict an electronic phase diagram of the entire range of rare earth monopnictides and monochalcogenides, 
composed of metallic, semiconducting and heavy fermion-like regions, and exhibiting valency transitions 
brought about by a complex interplay between ligand chemistry and lanthanide contraction. The calculations exploit the 
combined effect of a first-principles methodology, which can adequately describe the dual character of electrons, 
itinerant vs. localized, and high throughput computing made possible by the increasing available computational power. 
Our findings, including the predicted "intermediate valent" compounds SmO and TmSe, are in overall excellent 
agreement with the available experimental data. The accuracy of the approach, proven e.g. through the 
lattice parameters calculated to within $\sim$1.5\% of the experimental values, and its ability to describe localization 
phenomena in solids, makes it a competitive atomistic simulation approach in the search for and design of  
new materials with specific physical properties and possible technological applications.}

\end{abstract}

%\pacs{}
\maketitle

\section{Introduction}

Increased computer power, together with reliable electronic structure 
codes, have opened up the possibility of exploring electronic properties of huge numbers
of materials.\cite{Ortiz}  A systematic first principles based theoretical study of entire families of compounds has the potential of exploring 
new trends in their physical properties or discovering materials with improved characteristics of 
possible interest for technological applications. %such as new electron-phonon mediated superconductors. 

The ab-initio determination 
of materials properties requires electronic structure calculations that are 
accurate and parameter-free. With the advent of density functional theory (DFT),\cite{Hohenberg_Kohn}
quantum mechanical calculations of the electronic structure of solids were put on a rigorous theoretical
foundation. Nevertheless, the lack of knowledge of the exact exchange-correlation energy functional 
constitutes a substantial obstacle regarding the predictive power of the calculations, as different 
mechanisms are at work
depending on the relative importance of band formation energy on one hand and on-site electron-electron 
interaction and electron localization on the other. Standard electronic structure calculations, based on the local 
spin density approximation (LSDA)\cite{Kohn_Sham,Barth_Hedin} to density functional theory,
work very well for conventional materials like simple metals and their alloys, characterized by 
delocalized electron states and large band formation energies. However, for systems where localized electron
states occur, such as the 4$f$'s in rare earth compounds, the LSDA suffers from a sizeable
self-interaction (SI) error,\cite{Fermi_SI} 
brought about by the local approximation to the exchange-correlation
energy functional. Consequently, different assumptions and/or parameters need to be invoked to describe the
electronic structure of 
strongly correlated electron materials.\cite{Temmerman_Handbook}

The self-interaction corrected\cite{Perdew_Zunger} local-spin-density (SIC-LSD) energy functional is free of the SI 
error and fulfils the requirement of being parameter free and of sufficient accuracy to study 
trends in properties of entire families of materials containing both localized and itinerant 
electrons.\cite{Temmerman_LNP,Svane_QuantChem,Svane_SIC}
Here we apply the SIC-LSD to provide global understanding of the ground state electronic structure
of rare earth monopnictides and monochalcogenides and, through that, also demonstrate the predictive
power and accuracy of this approach.
Specifically, the present study covers %the following two families, the rare earth monopnictides
%and the monochalcogenides, where we distinguish 
140 distinct compounds, RX, with R referring to the rare earth atoms Ce, Pr, Nd, Pm, Sm, Eu,
Gd, Tb, Dy, Ho, Er, Tm, Yb, Lu, and X standing for the pnictide atoms N, P, As, Sb, Bi and 
the chalcogenide atoms O, S, Se, Te, Po. %The choice of the compounds for this case study has been 
%motivated by the fact that almost all of these materials are known to exist in nature. 
Most of these compounds are known to exist in nature, crystallizing predominantly
in the rocksalt structure.
Due to the intricate dual nature of the 4$f$-electrons, as a function of chemical environment these
compounds display an extraordinarily wide range of electronic, magnetic, optical, and magneto-optical properties with,
among other things, potential practical applications in the field of spintronics.\cite{Schmehl,LeClair} 
%enhanced nuclear cooling, 
%colossal magnetoresistance, non-linear optics and
%lasers.\cite{Karl,Verma} 
%In recent years there has been a lot of research towards understanding electronic 
%properties of these compounds, driven by the fact that some show interesting semiconducting properties
%for potential practical applications e. g. in the fields of electronics, non-linear optics and 
%lasers.\cite{Karl,Africa} 
A considerable number of theoretical studies on the RX compounds have resulted in greatly improved
insight into their properties.\cite{Duan,Petukhov,Larson,Antonov} 
However those calculations have often been guided by experiment, relying on parameters to reproduce
the observed properties.
The goal of the present systematic study 
%of the rare earth monopnictides and 
%monochalcogenides 
is to establish a paramater-free unified theoretical picture 
of the RX electronic properties and possibly provide directions for further experimental investigations.
With future applications in mind, 
we show how large scale computational infrastructures can potentially be
exploited in the 
search for the novel 
rare earth related materials that play an increasingly prominent role in innovative 
technologies and the quest for renewable energies.\cite{Ramsden}

\section{Theoretical Background}

In the LSD approximation the total energy functional is decomposed as
\begin{equation}
\label{ELSD}
E^{LSD}=\sum_{\alpha }^{occ.}\langle \psi_{\alpha} | \hat{T} | \psi_{\alpha} \rangle+U[n]+V_{ext}[n]+E_{xc}^{LSD}[n_{\uparrow},n_{\downarrow}], \\
\end{equation}
where $ \hat{T}$ is the kinetic energy operator, and $V_{ext}[n]$ represents the interaction with an external potential. The electron-electron interaction
is given as the sum of the Hartree term $U[n]$ and the local exchange-correlation term, $E_{xc}^{LSD}[n_{\uparrow},n_{\downarrow}]$.

In the Hartree-Fock approximation, the expectation value of the electron repulsion operator 
in some $f^n$ multiplet state $\Psi^n(LSJ)$ may be written as~\cite{Jorgensen, Nugent}
\begin{equation}
\label{Hartree}
\langle \Psi^n(LSJ) \mid \sum \frac{1}{r_{ij}} \mid \Psi^n(LSJ) \rangle = \frac{1}{2} n (n-1) E^0 +\alpha_1 E^1 + \alpha_2 E^2+\alpha_3 E^3,\\
\end{equation}
where $\Psi^n(LSJ)$ is in general given by a linear combination of $n\times n$ Slater determinants.
Since $\sum \frac{1}{r_{ij}}$ commutes with total $\bf L$, $\bf S$, and $\bf J$, the above matrix elements are equal for $M_J$ quantum numbers
belonging to one ($LSJ$) family. $E^i$ are the Racah parameters,~\cite{Racah} which are related to the Slater two-electron integrals (defined later). The coefficients
$\alpha_i$ depend on the particular $LSJ$-configuration chosen.

The quantity $E^0$ is the Coulomb repulsion between two $f$-electrons, and the Coulomb energy term in the LSD total energy functional (\ref{ELSD})
corresponds to the $\frac{1}{2}E^0n^2$ part of the first term in Eq. (\ref{Hartree}). 
The term, $-\frac{1}{2}E^0n$, in Eq. (\ref{Hartree}), cancels an equivalent contribution in the Hartree energy term, 
representing the interaction of an electron with itself.
A similar cancellation does not occur in the LSD functional, which therefore carries an unphysical self-interaction.~\cite{Perdew_Zunger}
This interaction, which tends to be insignificant for extended band state, 
may lead to uncontrollable
errors in the description of atomic-like localized states, for example the $f$-electrons in the rare earths.
Following the suggestion by Perdew and Zunger~\cite{Perdew_Zunger}, in the SIC-LSD methodology, the LSD 
functional (1) is corrected for this spurious
self-interaction
by adding an explicit energy contribution for an electron to localize.
The resulting, orbital dependent, SIC-LSD total energy functional has the form~\cite{Temmerman_LNP,Svane_SIC}
\begin{equation}
\label{Esic}
E^{SIC-LSD}=E^{LSD}+E_{so}-\Delta E_{sic},\\
\end{equation}
where
\begin{eqnarray}
\label{Esicdetail}
\hspace{-4mm}
\Delta E_{sic}&=&\sum_{\alpha }^{occ.}\delta _{\alpha }^{SIC}=\sum_{\alpha }^{occ.} \left \{ U[n_{\alpha}]+E_{xc}^{LSD}[\bar{n}_{\alpha}] \right \},\\
E_{so}&=&\sum_{\alpha }^{occ.}\langle \psi_{\alpha} | \xi(\vec{r})\vec{l}\cdot\vec{s} | \psi_{\alpha} \rangle.
\end{eqnarray}
The self-interaction energy (4) consists of the self-Coulomb and self-exchange-correlation energies of the occupied orbitals $\psi_{\alpha}$
with charge density $n_{\alpha}$ and spin density $\bar{n}_{\alpha}=(n_{\alpha}^{\uparrow},n_{\alpha}^{\downarrow})$.
For itinerant states, the self-interaction $\delta _{\alpha }^{SIC}$ vanishes identically, while
for localized (atomic-like) states $\delta _{\alpha }^{SIC}$  may be appreciable.
Thus, the self-interaction correction constitutes a negative energy
contribution gained by an electron upon localization,
which competes with the band formation energy gained by the electron
if allowed to delocalize and hybridize with the available conduction states.
Different localized/delocalized configurations are realized by assuming different numbers of localized
states - here $f$-states on rare earth atom sites.
The SIC-LSD approach is fully ab-initio,
as both localized and delocalized states are expanded in the same set of basis functions,
and are thus treated on an equal footing.
If no localized states are assumed, $E^{SIC-LSD}$ coincides with the
conventional LSD functional, {\it i.e.}, the Kohn-Sham minimum of the $E^{LSD}$
functional is also a local minimum of $E^{SIC-LSD}$.

The spin-orbit interaction (5) couples the band Hamiltonian for the spin-up and spin-down channels,
{\it i.e.}  a double secular problem must be solved.
The spin-orbit parameter,
\[
\xi(r)=-\frac{2}{c^2}\frac{dV}{dr},
\]
 in atomic Rydberg units,
is calculated from the self-consistent potential $V$.

Of the remaining terms in Eq. (\ref{Hartree}), the $E^1$ term accounts for Hund's first rule.
The $E^2$ term only contributes to the level spacing of excited multiplets and
will not be relevant in a functional for the ground state energy. The $E^3$ term, describes the level spacing between 
multiplets in the maximum spin configuration, i.e. this term accounts for Hund's second rule, by which 
the multiplet of maximum total orbital momentum, $L$, has the lowest energy. 
With respect to the average $f^n$ (maximum spin)
energy (the grand bary center), the maximum $L$ multiplet is lowered by $-jE^3$,
where $j=(0,9,21,21,9,0,0)$ for $n=(1-7)$, and $n=(8-14)$ respectively.
It is an atomic effect
which results in an increased stability at 1/4 and 3/4 filling of the 4$f$-shell, and this second Hunds's rule effect is
often referred to as the tetrad effect (TE).~\cite{Nugent}
Although, both LSD and SIC-LSD take into account Hund's first and third
rules, with respectively the exchange interaction and spin-orbit coupling included in the total energy functional,
Hund's second rule is not defined for the homogeneous electron gas, underlying LDA.\cite{Eriksson_OP}
To account for it in our total energy calculations, we add (a posteriori) to the SIC-LSD total energy functional
the relevant correction 
\begin{equation}
\label{tetradE}
\Delta E_t=-jE^3.\\
\end{equation}

The $E^3$ parameter is equivalent to Racah's $B$-parameter,\cite{Racah} and is given in terms of the
reduced Slater integrals $F_{k}$ as:
\begin{equation}
 E^{3}=\frac{1}{3}(5F_{2}+6F_{4}-91F_{6}),
\end{equation}
where $F_k=F^k/D^k$. Here $D^k$ are numerical constants ($D^k$=225, 1089, 7361.64 for $k$=2, 4, 6) and
the Slater integral $F^k$ is defined through
\begin{equation}
F^k=e^2\int_0^{\infty} \int_0^{\infty}\frac{(r_<)^k}{(r_>)^{k+1}}\;\phi_{4f}^2(r_1)\phi_{4f}^2(r_2)\; r_1^2dr_1\; r_2^2dr_2.
\end{equation}
Here, $\phi_{4f}$ is the f-partial wave as calculated in the self-consistent crystal potential, and $r_<$  ($r_>$) denotes the smaller (larger) 
of the variables $r_1$ and $r_2$. 
In reality, correlation effects
in the solid environment tend to reduce the multiplet energy level splittings.
Hence our calculated TE values
are on average 15-20\% larger than their experimental counterparts.\cite{Nugent}

%Hund's second rule refers to the multiplet structures and
%states that the multiplet of maximum total orbital momentum has the lowest energy. It is an atomic effect 
%which results in an increased stability at 1/4, 1/2, 3/4 and 4/4 filling of the 4$f$-shell, and hence this second Hunds's rule effect is
%often referred to as the tetrad effect (TE).~\cite{Jorgensen,Nugent} To account for the latter, in the SIC-LSD approach we add 
%(a posteori) to the total energy functional the relevant energy correction,
%in terms of the associated Slater integrals. Our calculated TE values
%This gives rise to a more complete approach which, without relying on any experimental input, 
%can scan entire classes of compounds and, through a systematic analysis of their ground state electronic 
%%structures, provides a reliable tool for the search of materials with potentially interesting 
%properties for technological applications. 
\section{Calculational details}
Given the total energy functional $E^{SIC-LSD}$, the computational procedure is as for the LSD case,
{\it i.e.} minimization is accomplished by iteration until self-consistency.
In the present work, the electron wavefunctions are expanded in
the linear-muffin-tin-orbital (LMTO) basis functions.\cite{Andersen_LMTO}
The atomic spheres approximation (ASA) is used, whereby the crystal volume is divided into
slightly overlapping atom-centered spheres of a total volume equal to the actual volume.
A known shortcoming of the ASA is that different crystal structures have different degrees of overlap of the ASA
spheres resulting in substantial {\it relative} errors in the evaluation of the total energy. While this inhibits
the comparison of energies of different crystal structures, when comparing
the energies of different localization/delocalization scenarios within the same crystal
structure the ASA error is of minor influence. To improve the packing of the structure
empty spheres have been introduced on high symmetry interstitial sites.
Two uncoupled energy panels have been
considered when constructing the LMTO's to ensure an accurate description of semicore states.
The $LS$-coupling scheme is adopted for the localized $f$-states by starting the iterations with
Wannier states of appropriate symmetry. During iteration to self-consistency the symmetry of the Wannier states may change, however
grossly retaining their overall characteristics due to the fact that the energy scale of spin-orbit interaction is smaller than that of exchange,
but larger than that of crystal field for the $f$-states.
The rocksalt structure with the magnetic moments arranged ferromagnetically was assumed in all the calculations.
To conduct the present large scale systematic study of rare earth compounds, in addition to a reliable theory and computer codes,
we have made an extensive use of grid computing and automated data/metadata management 
tools.\cite{Tyer_RCommands,Calleja} % developed by the eMinerals project\cite{Dove,Calleja}.
%This allowed us to perform many thousands of discrete calculations, distributed over the available grid 
%infrastructure, to determine the corresponding electronic structures, global energy minima and lattice parameters,
%and from those a unified electronic phase diagram of all the rare earth monopnictides and monochalcogenides. 

\section{Total energies and valency transitions}

\begin{figure}
\begin{center}
\includegraphics[width=220mm,clip,angle=0,scale=0.60]{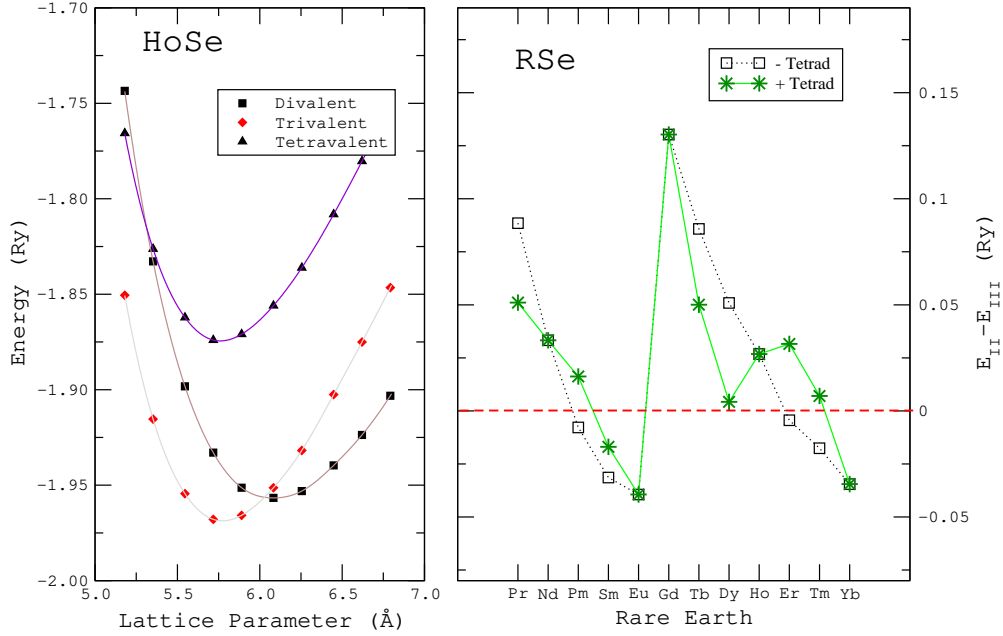}\\
\caption{
\label{XSe} a) Total energy as a function of lattice parameter for the divalent, trivalent and tetravalent 
configurations of the Ho ion in HoSe. b) The energy differences between the divalent and trivalent 
configurations, $\Delta E_{II-III}$, for the rare earth selenide series (RSe). Here the squares refer to 
the calculations without the tetrad effect, while the stars give the results with the tetrad effect included.
}
\end{center}
\end{figure}

%As already mentioned, the SIC-LSD methodology is based on the energy minimization with respect to 
In the SIC-LSD methodology, both the itinerant and
localized limits are described by the same energy functional, which enables us to
predict the ground state of correlated electron materials from total energy considerations.\cite{Temmerman_LNP,Svane_QuantChem}
Hence, one can
investigate localization phenomena in solids\cite{Temmerman_Handbook}
by realizing and studying different localization/delocalization scenarios, giving rise to different valency
configurations, and through total energy minimization determine the ground state electronic structure and
valency configurations of the compound under
consideration.\cite{Petit_felectron} This nominal valency
is defined as an integer number of electrons available for band formation, 
\begin{equation}
N_{val}=Z-N_{core}-N_{SIC},
\end{equation}
obtained by subtracting from
the atomic number ($Z$) the sum of core ($N_{core}$), and localized  ($N_{SIC}$) electrons.
In this paper we will be using two interchangeable nomenclatures, $f^n$ and R$^{m+}$, to describe the configuration of the
rare earth ion, implying $n=N_{SIC}$ and $m=N_{val}$, respectively. The total number of $f$-electrons may be larger
than $n$, because, in addition to the
$n$ localized $f$-states, the band states also contribute to the total $f$-count on a given ion. Note that our calculated valencies, refer
to the number of rare earth electrons that contribute to bonding, and thus do
not necessarily coincide with the nominal (ionic) valency of a compound. 

In our study of the rare earth monopnictides and monochalcogenides,  
for each given compound the self-consistent calculations involve the total energy minimization as a function
of lattice parameter 
with respect to
a number of different possible localization/delocalization scenarios,
in order to determine the global energy minimum and 
the corresponding equilibrium volume. To understand what is involved in this 
process, we explain the minimization procedure on the example of HoSe depicted in Fig. \ref{XSe}a.
What is seen in the figure are total energy curves, as a function of lattice parameter, for three
energetically relevant valency configurations. The way they come about is based on electronic structure considerations.
The Ho atom contributes 67 electrons to the total electron count of HoSe, 54 of which are considered to be %either 
part of the core (and semicore), and thus not contributing signifcantly to its chemistry. The remaining 13 electrons on the 
outer most shells, 11 of which are the $f$-electrons, are chemically active and constitute the valence electrons. 
The latter can either hybridize with the other band electrons or localize on site, playing no role 
in band formation. Assuming all the 11 $f$-electrons to be localized, means that only two $s$-electrons of Ho
contribute to band formation, a scenario which we refer to as divalent and which represents, with 
respect to $f$-electrons, the fully localized limit. 
Configurations with one or two delocalized (bonding) $f$-electrons 
are respectively referred to as trivalent and tetravalent. 
%, as seen in Fig. \ref{XSe}a. 
The LDA scenario, which treats all the ligand and rare earth valence electrons, including the 11 
$f$-electrons,  as band electrons, is found energetically unfavourable by $\sim$ 1 Ry, as compared to 
any of the other scenarios, and is therefore not shown in Fig. \ref{XSe}a. Altogether at least some forty 
self-consistent calculations
are required to establish the global energy minimum for a compound like HoSe which, as can be seen in Fig. \ref{XSe}a, 
is obtained in the trivalent scenario. 

From the calculated local energy minima we can evaluate the energy differences between respectively 
the divalent and trivalent configurations,
$\Delta E_{II-III}$, and the tetravalent and trivalent configurations, $\Delta E_{IV-III}$,
both of which are a measure of stability of the ground state configuration.
From Fig. \ref{XSe}a, we find for HoSe $\Delta E_{IV-III}$= 90 mRy and $\Delta E_{II-III}$= 20 mRy, indicating
a rather stable trivalent ground state, with no tendency towards delocalization of an additional 
$f$-electron, but with some, albeit marginal, tendency to localization of an additional electron. 
Thus whilst we predict a trivalent ground state at ambient conditions, with respect to possible phase 
transformations, for example under negative pressure,
$\Delta E_{II-III}$ is the relevant energy scale to consider in HoSe.
%However, at ambient conditions we predict a trivalent ground state, with the corresponding calculated 
%equilibrium lattice parameter equal to 5.55 \AA. 
The calculated trivalent equilibrium lattice parameter is equal to 5.55 \AA, differing
from the divalent minimum by about 5\%. 
%which is a substantial difference, 
%indicating that by comparing the calculated and experimental
%lattice parameters, one can easily establish whether the correct ground state has been predicted.

In Fig. \ref{XSe}b,
we present the calculated total energy differences, $\Delta E_{II-III}$, 
for the entire rare earth selenide series (RSe) from Pr onwards. 
%ption of CeSe, where the divalent state is 
%energetically unfavourable, and instead the tetravalent and trivalent solutions provide the relevant energy scale. %is presented. 
The two curves shown refer respectively to 
the SIC-LSD calculations with (stars) and without (squares) the TE.
The positive values indicate a trivalent ground state, whereas the negative values refer to the divalent 
configuration being energetically more favourable. One can see that without the TE, 
$\Delta E_{II-III}$ decreases monotonously from trivalent PrSe to divalent 
EuSe, and again from trivalent GdSe to divalent YbSe. The increasing trend towards divalency is associated 
with the enhanced localized nature of the $f$-electrons and can be understood as a manifestation
of the well known lanthanide contraction. A small $\Delta E_{II-III}$ indicates that for the given 
compound no clear ground state emerges, as the divalent and trivalent configurations 
are close to energetically degenerate, due to the competition
between band formation and self-interaction correction (localization) energies.

The overall trend of decreasing energy differences is still recognizable in both halves of the selenide series  
when the TE is included. However, due to the abrupt decrease in $\Delta E_{II-III}$ at DySe
a characteristic hump, also observed in experiments on rare earths\cite{Johansson_Rmetals,Patthey}, 
develops in the later part of the series, 
implying that this compound is considerably less trivalent
than one would expect from the SIC-LSD result prior to taking Hund's second rule into account.
Even more important are the changes in predicted ground state configurations for ErSe and TmSe 
(as well as PmSe), indicating that the tetrad effect can bring about a qualitative change 
in compounds where the original energy differences
are small.
Otherwise, inplementing Hund's second rule has only a quantitative effect on the $\Delta E_{II-III}$ of those
compounds that from the outset are either solidly trivalent or divalent. 

\begin{figure} 
\begin{center}
\includegraphics[width=230mm,clip,angle=0,scale=0.60]{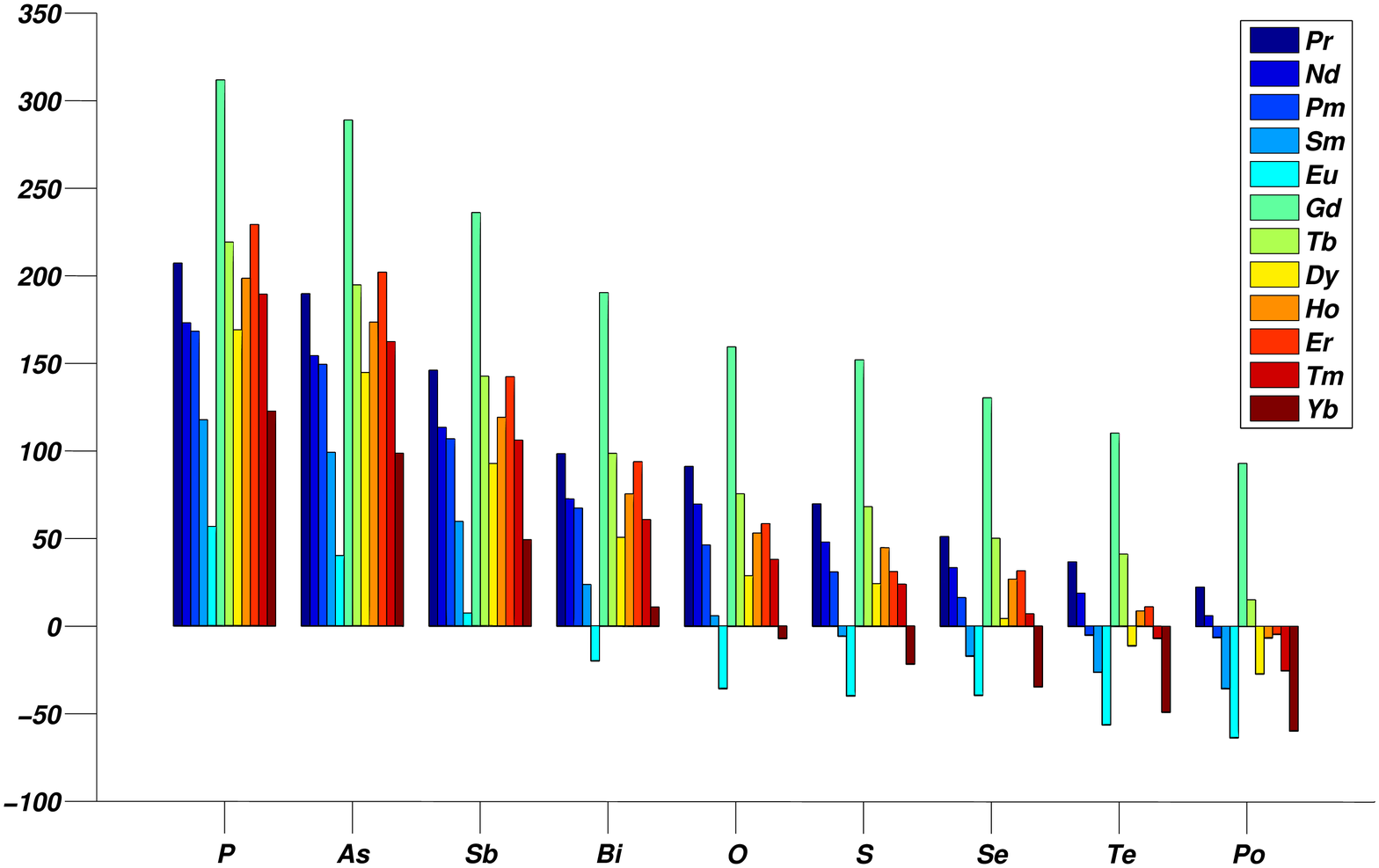}
\includegraphics[width=220mm,clip,angle=0,scale=0.65]{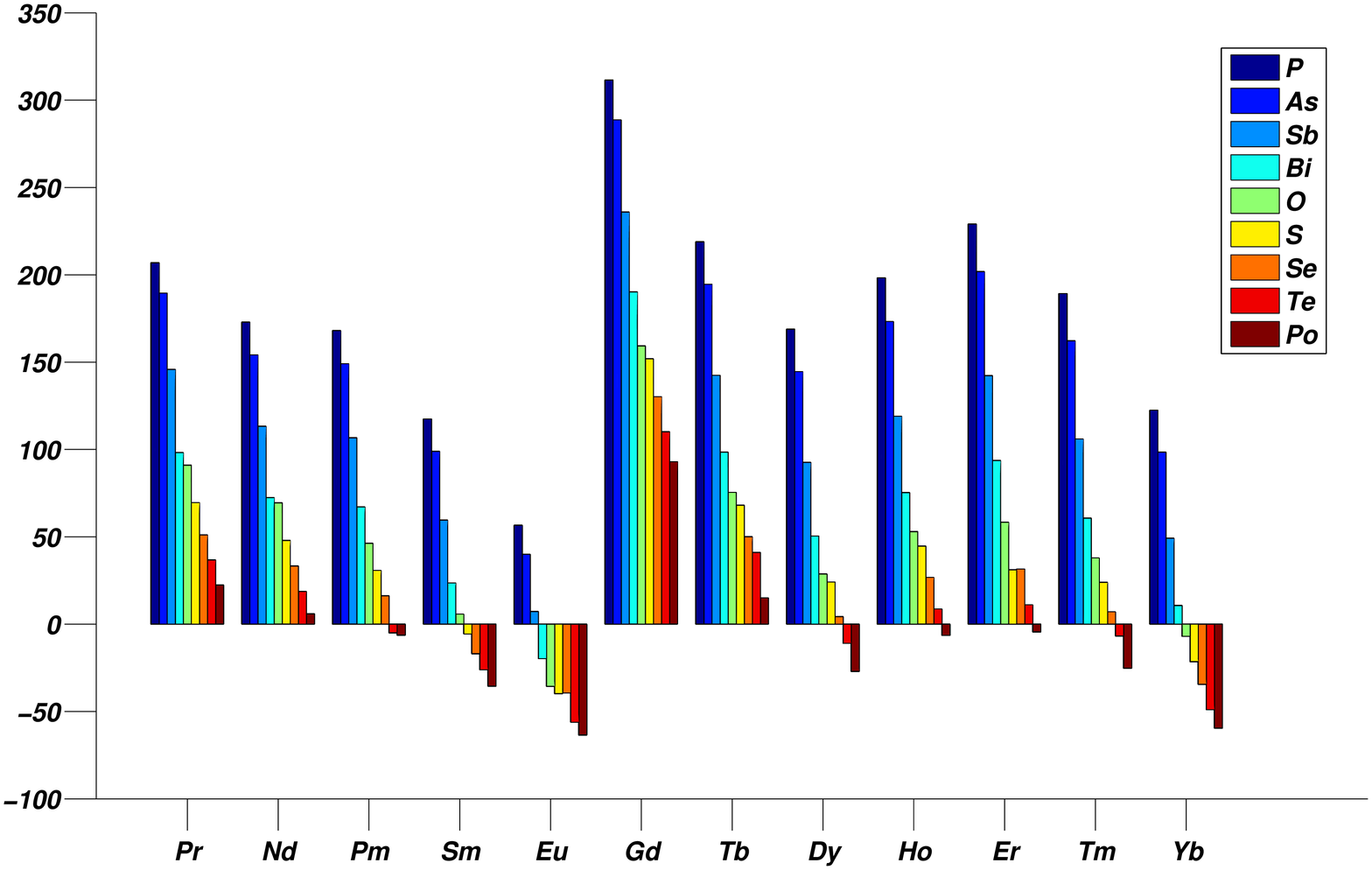}\\
\caption{
\label{total}
a) The energy difference $\Delta E_{II-III}$ as a function of ligand ion for all the rare earths.
b) The energy difference $\Delta E_{II-III}$ as a function of rare earth ion for all the ligands.
}
\end{center}
\end{figure} 

Taking the TE into account, the divalent-trivalent energy differences, $\Delta E_{II-III}$,
for all the RX compounds (except for the rare earth nitrides and the Ce compounds) are
collected in Fig. \ref{total}. 
The positive and negative values again indicate respectively
a trivalent and divalent ground state configurations.
In Fig. \ref{total}a, for each of the ligand ions the $\Delta E_{II-III}$ data are plotted as clusters of bars, representing 
the rare earth ions ranging from Pr to Yb. 
We notice that the total energy trends, observed for the selenide series in Fig. \ref{XSe}b,
are reproduced in all the ligands, but overall, in the chalcogenides the 
divalent configuration becomes increasingly more favourable as a function of ligand size, whilst in the pnictides 
the trivalent configuration remains the ground state, even for the
"late" rare earth Eu and Yb (with the exception of EuBi).
The trivalent to divalent localization transition thus mainly pertains to the rare earth chalcogenides.
In Fig. \ref{total}b, the same total energy data are depicted as a function of rare earth, 
with the ligands plotted as clusters of bars ranging from P to Po for each of the rare earth ions.
In this representation one can clearly see the combined effect of the lanthanide contraction, leading gradually towards 
the more preferred
divalent configurations, and the repetition of this trend due to the half-filled shell effect. 
%Inspecting Fig. \ref{XSe}c, one can see that overall the trivalent to divalent transition,
%associated with localization of one additional $f$ electron, occurs gradually
%with increasing atomic number of rare earth ion, from Pr to Eu, and again from Gd to Yb, reflecting
%the lanthanide contraction. The jump seen between Eu and Gd arises due to half-filling of the 4$f$ shell.
%For a given rare earth the trend is that of decreasing trivalency from N to Bi and from O to Po,
%namely as a function of increasing ligand size.
%Overall, trivalency is the dominant ground state configuration,
%with divalency being observed mostly in the chalcogenides of close to half-filled and fully-filled rare earths.
%The more divalent character of Dy compounds is again related to the tetrad effect.

Although $\Delta E_{II-III}$
is the relevant energy scale for the large majority of the RX, 
there are a number of exceptions where the delocalized tetravalent scenario
is energetically more favourable than the localized divalent scenario 
even though according to our calculations it never actually 
becomes the true ground state. The 
trivalent-tetravalent energy difference, $\Delta E_{IV-III}$, is the more relevant energy scale 
for the Ce compounds for example, as the divalent configuration
is always very unfavourable, and in most cases could not even be converged. 
In the case of CeN, and to some degree in CeO, the tetravalent and trivalent scenarios are actually 
close to energetically degenerate.  Since the divalent scenario is so highly unfavourable, the Ce 
pnictides and chalcogenides, as well as all the other rare earth nitrides, are not shown in Fig. \ref{total}. 
For the LuX compounds $\Delta E_{II-III}$ is not adequate, as a divalent scenario
would here imply 15 localized $f$-electrons.
\begin{figure}
\begin{center}
\includegraphics[width=160mm,clip,angle=0,scale=1.00]{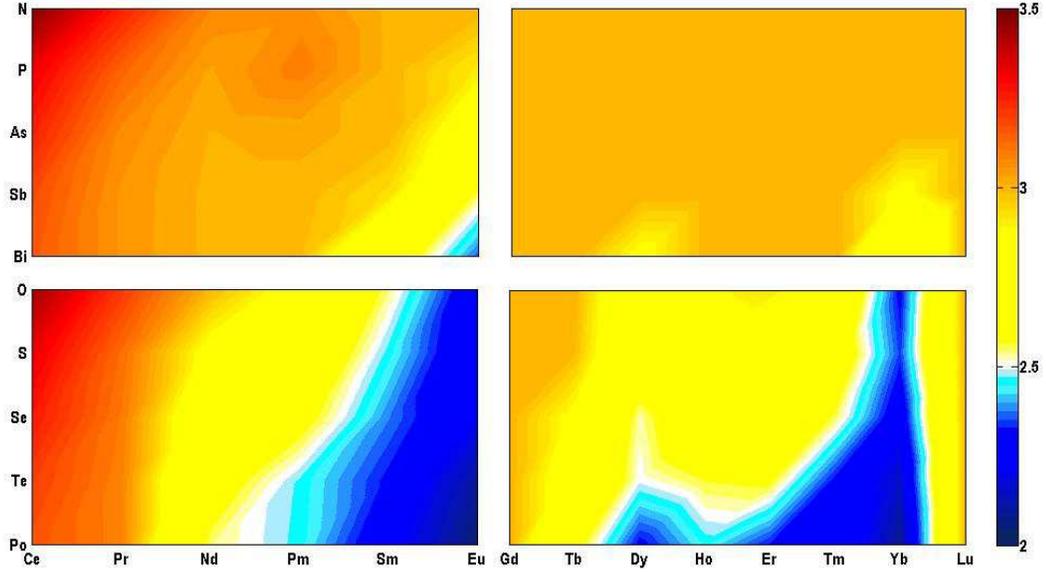}
\caption{
\label{Etot} 
%"valency phase diagram" 
The valency phase diagram for the entire manifold of rare earth monopnictides and monochalcogenides.
The blue areas mark divalent compounds, while trivalency is indicated by the yellowish colour.
The red parts of the diagram refer to the close to tetravalent compounds (mostly Ce compounds).
The light blue/white regions represents the border line cases between divalent and trivalent
configurations. The colour scheme refers to the valencies, obtained by scaling the calculated energy
differences such that a value of 3.5 indicates trivalent-tetravalent degeneracy as observed in CeN,
whilst values smaller than 2.5 refer to divalent ground state configurations.
}
\end{center} 
\end{figure} 

%%%%%%%%%%%%%%%%%%%%%%%%%%%%%%%%%%%%%%%%%%%%%%%%%%%%%%%%%%%%%%%%%%%%%%%%%%%
%A unified picture emerges for all the rare earth monopnictides and monochalcogenides when we focus 
%on the energy differences between those two configurations 
%that are closest in energy, i.e. either $E_{II}-E_{III}$ or $E_{IV}-E_{III}$, depending on which is more 
%relevant for a given compound. The outcome is presented as a "valency phase diagram" shown in Fig. \ref{Etot}. 
%Here the yellow areas indicate that the trivalent configuration is strongly preferred, which is the case 
%for a large number of compounds, especially the pnictides.  The redish areas indicate
%that the tetravalent configuration is becoming more competitive, which as we stated earlier, concerns mostly the CeX and early rare earth nitrides. Finally we have the blue islands indicating that here the divalent 
%configuration is stable, while the light-blue and whitish
%areas represent compounds that are situated at the boundary of the trivalent-divalent transition.  
%The colour scheme refers to the valencies, obtained by scaling the calculated energy differences such that a value of 3.5 indicates
%trivalent-tetravalent degeneracy as observed in CeN, whilst values smaller than 2 refer to divalent ground state configurations. 
%
%%%%%%%%%%%%%%%%%%%%%%%%%%%%%%%%%%%%%%%%%%%%%%%%%%%%%%%%%%%%%%%%%%%%%%%%%%%
As shown in Fig. \ref{Etot}, a unified picture for all the RX compounds emerges when we map the calculated energy differences into ground state
valencies. Here the valency scale is defined such that it runs from 3.5 for the tetravalent/trivalent degenerate 
compounds ($E_{IV}-E_{III}$=0) to 2.5 for
the trivalent/divalent degenerate ($E_{II}-E_{III}$=0) compounds. 
%For valencies between 3.5 and 2.5 the compounds are classified as trivalent, whilst
%valencies smaller than 2.5 indicate a divalent ground state. 
%on the energy differences between those two configurations that are closest in energy, i.e. either 
%$E_{II}-E_{III}$ or $E_{IV}-E_{III}$, depending on which is most relevant for a given compound.
%The outcome is presented as a "valency phase diagram" in Fig. \ref{Etot}, where the respective
%calculated energy differences have been translated into valencies, ranging from 3.5 for $E_{IV}-E_{III}$=0
%to 2.5 for $E_{II}-E_{III}$=0 
The yellowish 
areas, that dominate throughout the phase diagram, demonstrate the overall preferred trivalent 
ground state configuration for the large number of the RX compounds, in particular the monopnictides. 
The trend towards delocalizing an additional electron is only noticeable for the
CeX systems and some of the nitrides, as indicated by the more reddish areas of the phase diagram,
whilst the trivalent to divalent localization transition occurs around the regions indicated in
white. The 
blue zones represent the 
stable divalent configuration, which is the dominating ground state for many of the "heavy" chalcogenides, 
in combination with the rare earths situated in the middle and later parts of the series.

\section{Electronic structure}

\begin{figure}
\begin{center}
\includegraphics[width=150mm,clip,angle=0]{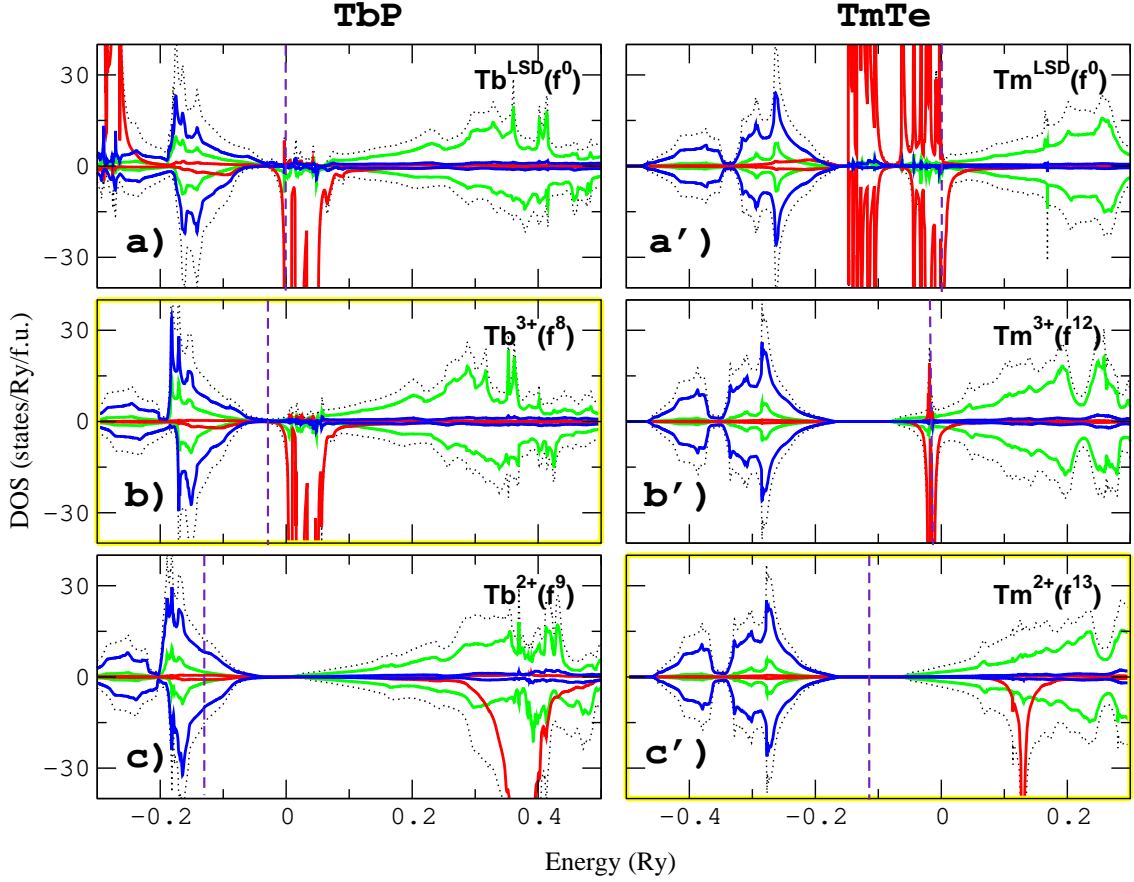}
\caption{
\label{dos} Left column: Density of states of TbP respectively for the a) LSD, b) trivalent and c) divalent Tb ion configuration.
Right column: Density of states of TmTe respectively for the a') LSD, b') trivalent, and c') divalent configuration of the Tm ion.
Density of states: total (dotted black), ligand $p$-projected (solid blue), rare earth $d$-projected (solid green), rare earth $f$-projected
(solid red). The Fermi level is indicated by the dashed vertical line. 
}
\end{center}
\end{figure}
\subsection{Density of states: chalcogen versus pnictogen chemistry}
The observed total energy trends can be understood in terms of variations in the ground state electronic structures brought about by 
changes in the atomic composition, the distance between atoms, and the relative position of the atomic energy levels.
In the RX compounds, the electronic structure results mainly from the $f$- and $d$-orbitals on 
the R-sites overlapping with the $p$-orbitals on the ligand-sites, as well as with the $f$- and $d$-orbitals on neighbouring R-sites.
In figure \ref{dos}, using TbP (column 1) and TmTe (column 2) as representative examples for the pnictide and chalcogenide
rare earth compounds, we depict the density of states (DOS) for three different valency scenarios, as realized respectively in 
the delocalized (Figs. \ref{dos}a and \ref{dos}a') , trivalent (Figs. \ref{dos}b and \ref{dos}b'),
and divalent (Figs. \ref{dos}c and \ref{dos}c') configurations. 
%Apart from the total DOS, the orbital decomposed DOS of those channels that are most relevant to the
%electronic structure at and around the Fermi level are shown, namely the Tb/Tm-ion $f$- and $d$-states and the pnictide/chalcogenide
%$p$-states. The position of the Fermi level is indicated by the vertical dashed line.

The LSD band picture, for both TbP (Fig. \ref{dos}a) and TmTe (\ref{dos}a'), is characterized by a large, $f$-derived, DOS at the
Fermi level. A noticeable difference is observed with respect to the relative strength of the exchange and spin-orbit interactions.
In TbP the exchange interaction dominates, which results in the splitting of the $f$-states into majority spin states, situated at the
bottom of the phosphorus $p$-band, and minority spin states, situated at the Fermi level. In TmTe the spin-orbit interaction has become 
more important, splitting the thulium $f$-states into $j=5/2$ and $j=7/2$ bands, with the exchange interaction being responsible for
the additional splitting within each of the two groups of bands. 

In spite of its interesting features, the band picture (Figs. \ref{dos}a and \ref{dos}a') is an unphysical representation of the RX compounds, 
as the large DOS at the Fermi level
is not observed experimentally. 
The trivalent scenarios 
of Figs. \ref{dos}b and \ref{dos}b' are obtained by localizing respectively eight $f$-electrons in TbP (Tb$^{3+}$ $\equiv$ Tb($f^8$)), 
and twelve $f$-electrons in TmTe (Tm$^{3+}$ $\equiv$ Tm($f^{12}$)). Compared to the LSD scenario the trivalent configuration
is found to be energetically more favourable by $\sim$ 700 mRy for TbP and $\sim$ 1200 mRy for TmTe.
In TbP the seven majority spin states , as well as one of the minority spin states, are 
no longer available for band formation and hybridization and have
vanished from the band picture in Fig. \ref{dos}b.
Similarly in the DOS for TmTe (Fig. \ref{dos}b'), the localization of the six $j=5/2$ $f$-electrons
as well as six of the $j=7/2$ electrons, implies that only two $f$-states are available for band formation. 
Notice that in the DOS plots only the band states are displayed, i.e. the itinerant valence states, including the delocalized 
$f$-states. The localized $f$'s are not shown since the SIC-LSD approach, which after all is a one-electron ground state theory,
does not give accurate removal energies of localized states due to electron-electron interaction effects,\cite{svane_DMFT} and the neglect of screening and relaxation
contributions.\cite{Temmerman_Pr} 

The trivalent DOSs, in respectively Figs. \ref{dos}b and \ref{dos}b', differ noticeably from each other, as in the former the Fermi level is situated 
in the gap between the filled phosphorus $p$-band and the unoccupied $fd$-band, whilst in the latter the Fermi energy is pinned to
the sharp $f$-peak above the conduction band minimum.
The P atom has three unoccupied $p$-orbitals and in the TbP crystal it can accommodate three electrons from the Tb ion through charge transfer
and hybridization which results in the filled $p$-band of Fig. \ref{dos}b. The Te atom on the other hand has one additional $p$-electron in its
outer shell, which is why the $p$-band in TmTe can only accommodate two electrons, and as a consequence of this, here in the trivalent Tm scenario, the
third electron is forced to occupy the $fd$-conduction band. 

The variation in $p$-orbital occupation is fundamental to the 
overall difference that we observe in figure \ref{Etot} with respect to the ground state valency configurations of the pnictides and chalcogenides.
Thus we find that when trying to localize an additional $f$-electron in TbP, the resulting divalent configuration, depicted in figure \ref{dos}c, is energetically
unfavourable compared to the trivalent scenario in figure \ref{dos}b by $\sim$ 250 mRy.
The $f^8$ to $f^9$ localization is
associated with charge transfer, and brings about the depopulation of the P $p$-states, as indicated by the Fermi level being situated in the $p$-band.
The removal of the bonding states results in a considerable loss in hybridization and Madelung energy, which is not compensated by the gain in  
self-interaction energy, and the divalent scenario will therefore not be realized in TbP.
In TmTe on the other hand, the localization transition from $f^{12}$ in Fig. \ref{dos}b' to $f^{13}$ in Fig. \ref{dos}c', is associated with a minor loss in band formation
energy, given the vanishing width of the occupied $f$-peak, and the gain in self-interaction (localization) energy is comparatively larger by
$\sim$ 35 mRy. Accordingly, the divalent scenario of Fig. \ref{dos}c', becomes the ground state scenario for TmTe.

\subsection{Electronegativity and lanthanide contraction}

Apart from the fundamental difference between the pnictide and chalcogenide electronic structure, the observed energy trends in each series,
 as a function of R-ion in figure \ref{total}a and as a function of ligand in figure \ref{total}b, are determined
by further factors such as ligand electronegativity, and lanthanide contraction.
\begin{figure}
\begin{center}
\includegraphics[width=160mm,clip,angle=0]{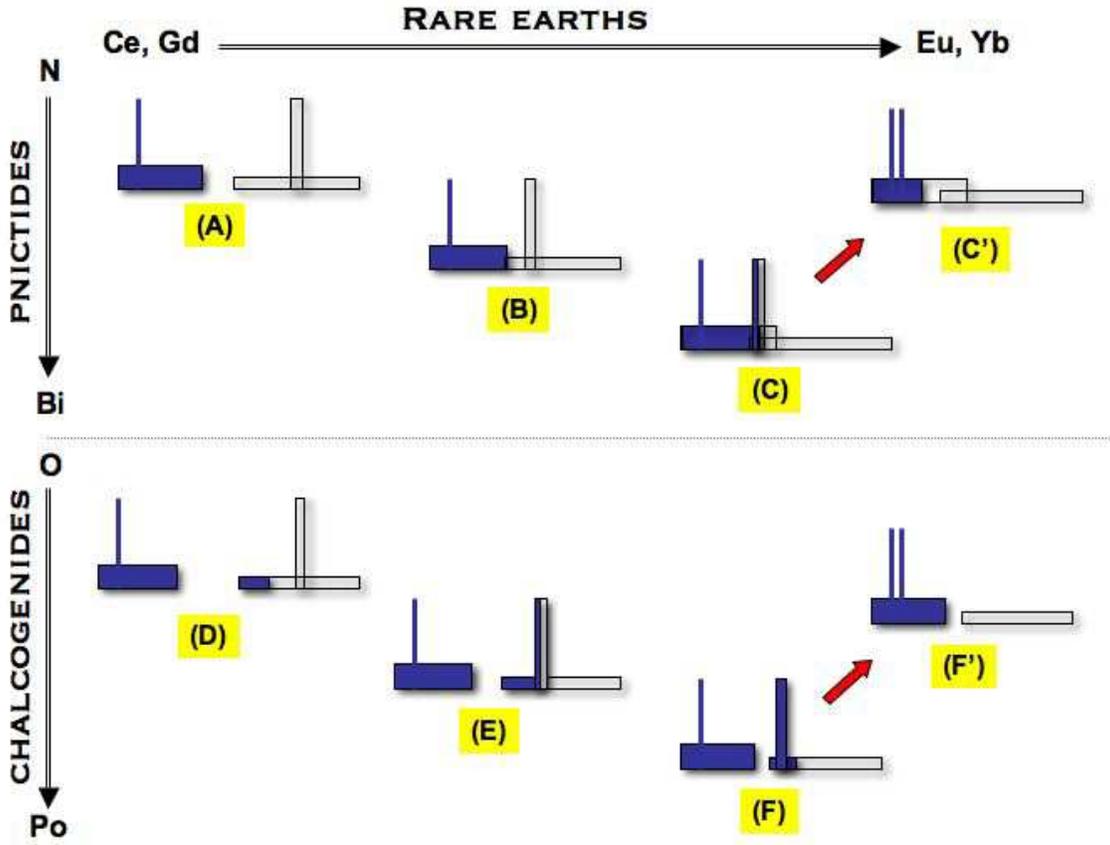}
\caption{Schematic representation of densities of states (DOS) for the rare earth monopnictides (upper panel) and
monochalcogenides (lower panel). Each DOS is composed of a number of rectangles representing
respectively the ligand $p$-states (situated lowest in energy), the rare earth $d$-states (situated highest in energy),  
and the occupied and unoccupied $f$-states (upright bars). Occupied states are shown in blue.
For each panel the three different scenarios ((A) to (C)) and ((D) to (F)), represent the different trivalent DOS resulting from the decrease in electronegativity
and the increase in $f$-state localization as one moves along the diagonal of the corresponding compound space.
Scenarios (C') and (F') refer to the divalent scenarios obtained respectively form (C) and (F) through
localizing an additional $f$-electron.
\label{schem}
}
\end{center}
\end{figure}
These changes in 
electronic structure are presented schematically in Fig. \ref{schem}. 
Here we depict the densities of states (DOS) for eight distinct scenarios observed throughout the rare earth compounds.
Each DOS is composed of several rectangles, referring respectively 
to the ligand $p$-states (situated lowest in energy), the rare-earth $d$-states, and the localized
and delocalized $f$-states.
The occupied states are indicated in blue.  
%Two main differences are observed between the pnictides 
%and chalcogenides. In the latter, due to the overall increased electronegativity, the $p$-states 
%are relatively lower in energy compared to the $d$-states, and more importantly, with the pnictide
%atom having one less occupied $p$-state compared to the chalcogenide atom, the corresponding conduction 
%band can accommodate three electrons from the rare earth through charge transfer and hybridization (scenario (A) in Fig. \ref{schem}). 
%A trivalent scenario is thus more readily realized in the pnictide compounds, as compared to the 
%chalcogenides where the third electron needs to be accommodated in the valence band (scenario (D)).
For both the pnictides and chalcogenides, the electronegativity decreases with increasing anion size, 
i.e. respectively from N to Bi and from O to Po. On the other hand, the $f$-electron localization 
increases due to the lanthanide contraction, as we move through the rare earth series from Ce to Eu, 
and again from Gd to Yb. In terms of DOS, these two trends manifest themselves respectively 
via ligand $p$-bands moving up in energy (towards lower binding energies), and unoccupied 
$f$-states moving towards higher binding energies.  In Fig. \ref{schem}, the combined effect of these 
two trends is depicted in the two sets of DOSs respectively along the diagonals of the pnictide ((A) $\rightarrow$ (C)) and chalcogenide 
compounds ((D) $\rightarrow$ (F)). As the $f$-states move towards lower energy they get pinned to the Fermi level and start 
filling up.  Ultimately the degree of $f$-band filling becomes such that the gain in energy associated 
with localizing the given $f$-state becomes more important than the corresponding loss in band 
formation energy, and the divalent scenario becomes the ground state (scenarios (C') and (F')).\cite{Temmerman_PhaseTrans} 
As seen in Fig. \ref{schem}, 
owing to the additional electron, in the chalcogenides the filling of $f$-states sets in much earlier
than in the pnictides, which explains why the trivalent scenario is relatively more dominant in the 
pnictides.

\section{Electronic Properties}

\subsection{Lattice parameters}

\begin{figure}
\begin{center}
\includegraphics[width=60mm,clip,angle=-90]{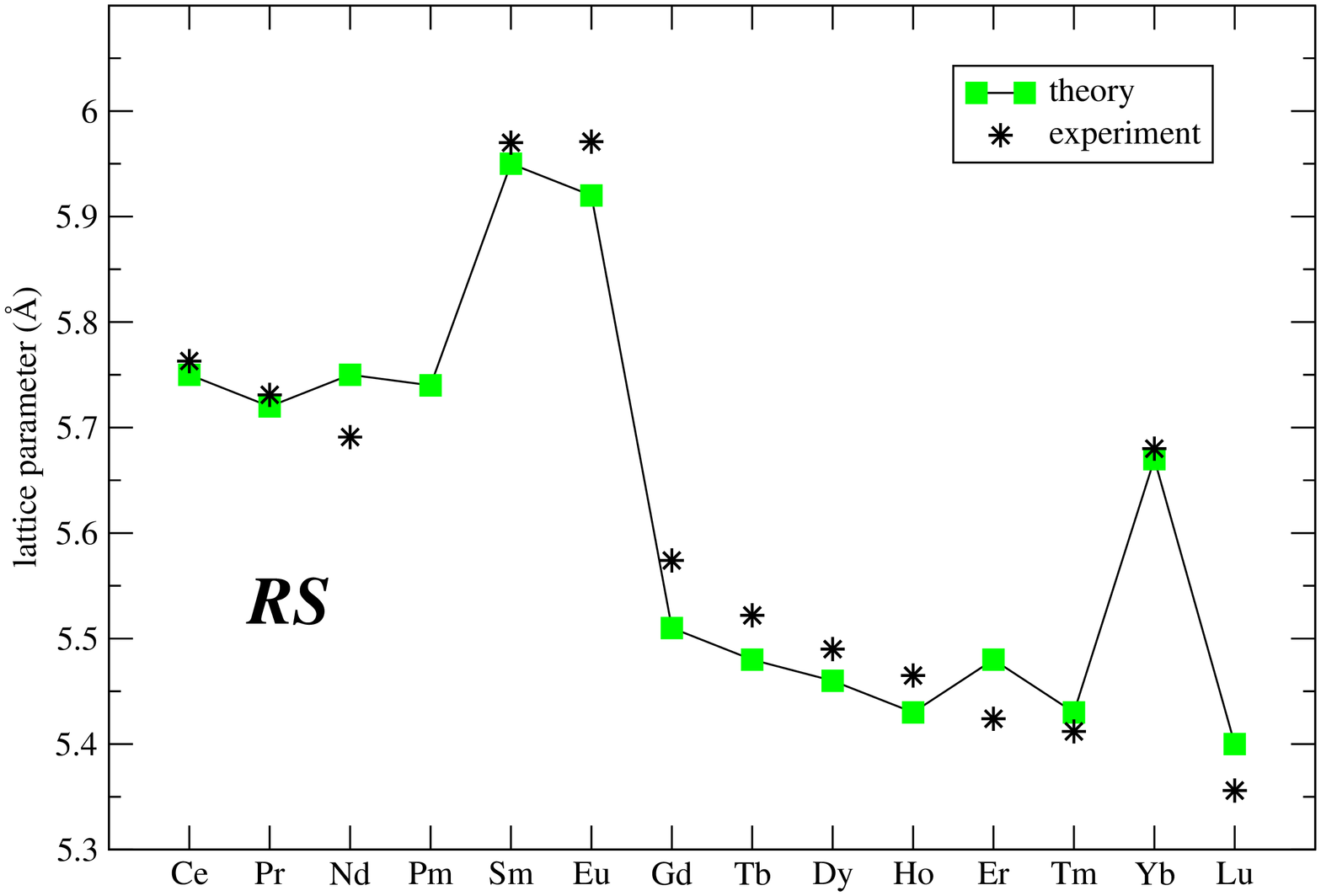}
\includegraphics[width=60mm,clip,angle=-90]{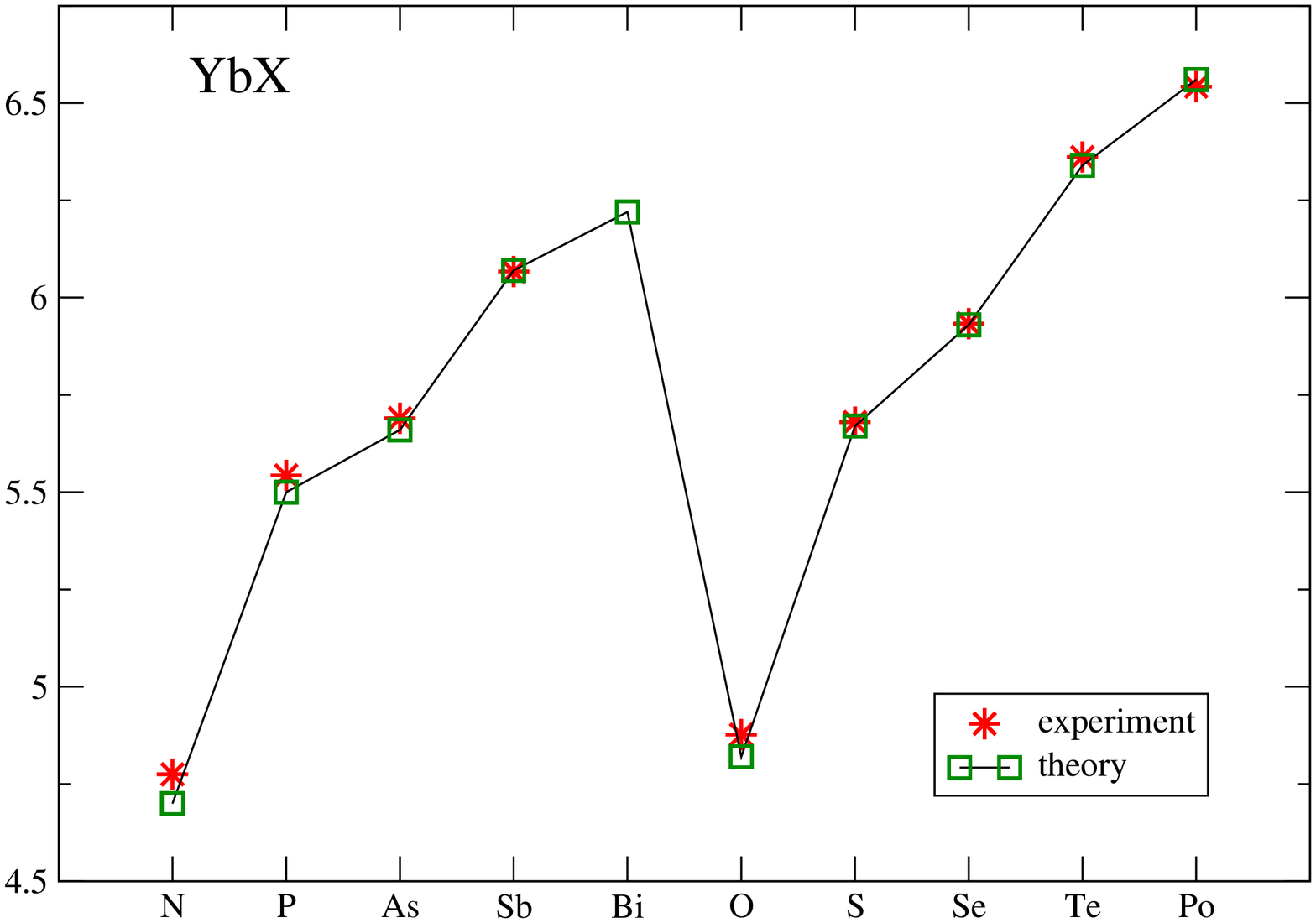}
\caption{
\label{alatAll} Lattice parameters of a) the lanthanide sulphides (RS) and b) the Yb pninctides and chalcogenides (YbX) 
as calculated in their ground state configuration.
}
\end{center}
\end{figure}

For each given compound, the calculated global energy
minimum
determines the equilibrium lattice parameter as well as the ground state. 
The calculated lattice parameters for the rare earth sulphides and the ytterbium pnictides and chalcogenides
are compared to their experimental counterparts
in respectively Figs. \ref{alatAll}a and \ref{alatAll}b. The agreement is good, reproducing the sudden increase in lattice parameter
at SmS, EuS and YbS, due to the increased $f$-electron localization.
We have observed that
quite systematically the calculated lattice parameters for the divalent and trivalent scenarios of a given compound 
differ by about 5 \%.
Therefore, a comparison to the experimental lattice parameters is expected to provide a
good indication as to whether the calculations predict the correct ground state and valency configuration.
\begin{figure}
\begin{center}
\includegraphics[width=160mm,clip,angle=0]{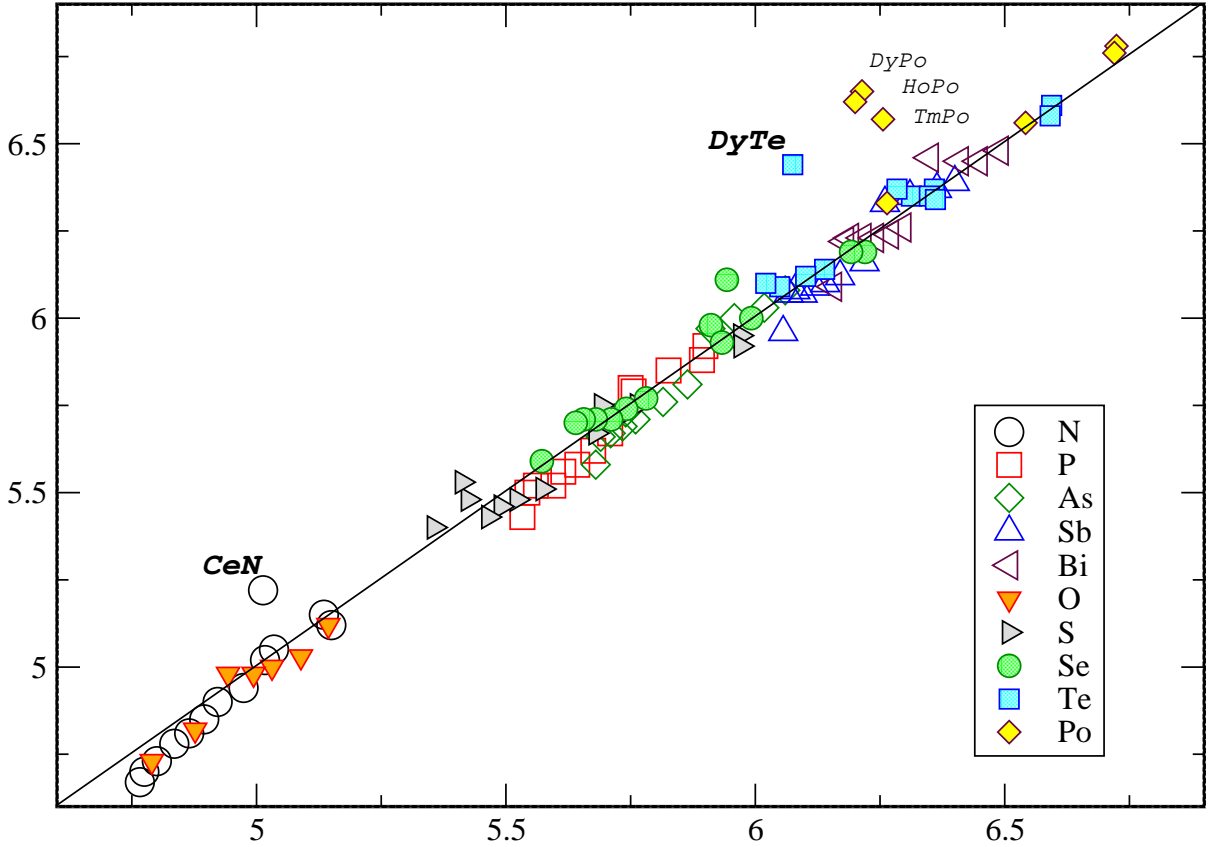}
\caption{
\label{alat} Lattice parameters of all the studied RX compounds: experimental (x-axis) vs. theoretical (y-axis) values.
}
\end{center}
\end{figure}
In Fig. \ref{alat}, the experimental equilibrium lattice
parameters for all the rare earth monopnictides and monochalcogenides\cite{Pearson} are plotted along the x-axis, whilst our calculated equilibrium 
counterparts are plotted along the y-axis.
%In figure \ref{alat} the experimental data are plotted along the x-axis whilst the calculated ground state values 
%are plotted along the y-axis.
In this representation, the proximity of the data points to the $x=y$ line is a measure of the agreement between 
theory and experiment. We find that for the large majority of the compounds the calculated equilibrium lattice parameters lie within 
$\sim$1.5\% of the experimental values, i.e. well below the 5\% that would indicate a wrongly predicted ground state 
configuration. The only exceptions are CeN, DyTe, and three of the polonide compounds. With respect to the 
latter compounds, polonium is difficult to handle, and it is unclear to what degree the
experimental results refer to stoichiometric samples.\cite{Kershner} 

It is well established from rare earth halide data\cite{Johnson2006} as well as XPS measurements on the rare earth elements\cite{Johansson_Rmetals}
that the divalent configuration becomes more competitive in Dy compounds due to the 3/4 filling of the $f$-shell.
The inclusion of the TE leads to a considerable reduction of $E_{II}-E_{III}$ for DyX compounds,
as can be seen on the example of DySe in Fig. \ref{XSe}b, and quite generally gives rise to the negative
dip around Dy in Fig. \ref{total}b. This trend reproduces overall very well the experimentally
observed data,\cite{Patthey} however we wrongly predict a Dy$^{2+}$ ground state in DyTe. This seems to indicate 
that in some cases we tend to
overestimate the tetrad effect's influence as a result of approximating
the respective Slater integrals by the fully localized atomic limit instead of considering also corrections
due to the solid state environment (screening/hybridization).
For CeN, the experimental lattice parameter seems to suggest
a tetravalent ground state,\cite{Jayaraman_CeX} whereas our calculations predict 
the trivalent configuration to be marginally more favourable than the tetravalent configuration by $\sim$3 mRy, which
on the other hand could be indicative of an intermediate valency scenario implied by experiments.\cite{Baer_CeX}  

In Fig. \ref{alat}, the wrongly predicted ground state of DyTe shows up as noticeably off diagonal data point,
but at the same time it also highlights the excellent agreement obtained for all the other cases investigated with the present
approach.
% hilst these few wrongly predicted ground state configurations point to some degree of uncertainety in the 
%corresponding calculated
%total energies, in Fig. \ref{alat}, where these predictions show up as noticeably off diagonal data points, 
%they highlight the excellent agreement obtained for all the other cases investigated with the present 
%approach. 
Thus, the SIC-LSD methodology is capable of predicting, with considerable accuracy, the 
ground state valency and
lattice parameter of the rare earth monochalcogenides and monopnictides, and the associated ground state electronic
structures can be expected to make 
valid predictions concerning 
the other physical properties of these materials.

\subsection{Electronic Phase Diagram}

\begin{figure}
\begin{center}
\includegraphics[width=180mm,clip,angle=0]{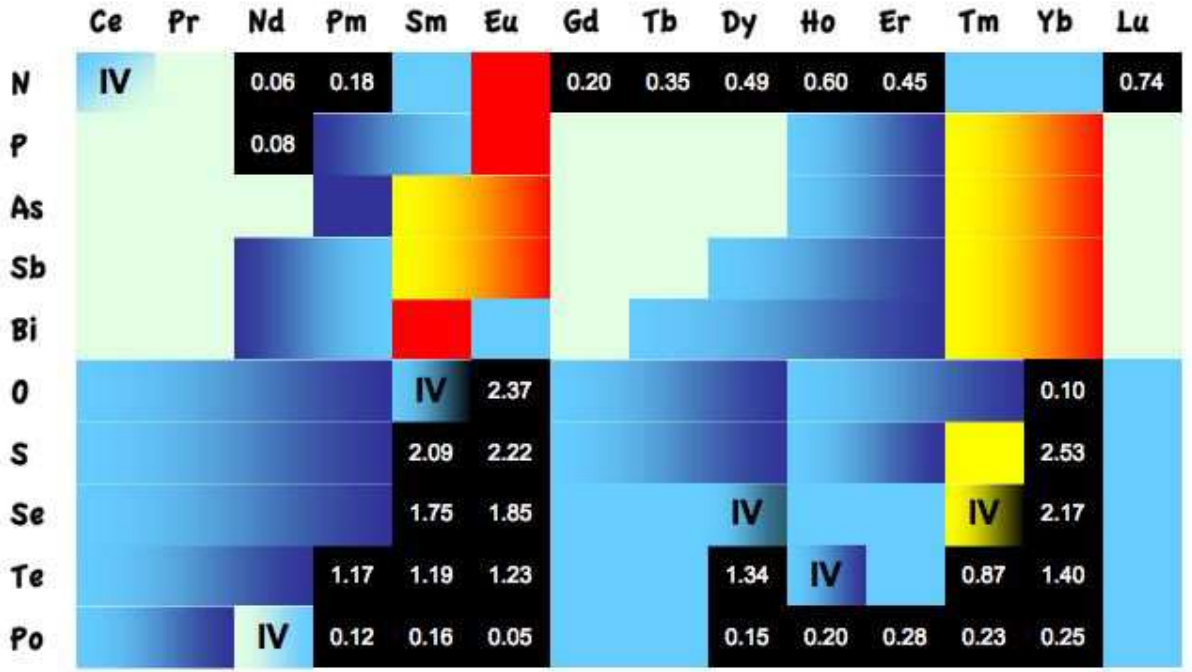}
\caption{A global picture of the electronic properties of all the studied pnictides and chalcogenides,
based on the densities of states at the Fermi energy E$_F$. The colour scheme is as follows: metals 
are marked in blue tones, with light blue for semimetals. Semiconductors are in black, whilst yellow to red is 
assigned to increasingly heavy fermion-like systems. The fields marked with "IV" indicate intermediate behaviour between
two different valency configurations.
\label{alat1}
}
\end{center}
\end{figure}

Here we concentrate on establishing global trends in the ground state electronic structure
which we base on the calculated DOS at the Fermi energy, E$_F$, as providing a guideline for 
predicting materials with specific properties when we systematically scan the full manifold of 
rare earth compounds. 
The corresponding results are summarized in the form
of an "electronic phase diagram" presented in Fig. \ref{alat1}, where based on the calculated DOS 
we distinguish between semiconductors, semi-metals, metals, and heavy fermion-like compounds. 

Looking at the phase diagram, we see that our calculations predict a considerable
number of compounds to be semiconductors. In particular, with respect to the pnictides, 
NdN, PmN and the majority of the rare earth nitrides from GdN
onwards are found to be small gap semiconductors,
with energy band gaps ranging from 0.06 eV in NdN to 0.6 eV in HoN. 
Few conclusive conductivity measurements for the nitrides exist, 
and it is for example still not fully established whether GdN
is semiconducting\cite{Granville} or semi-metallic.\cite{Leuenberger}
Our calculations predict GdN to be a semiconductor, although
in an earlier SIC-LSD calculation, where spin-orbit coupling was not taken into account, 
a half-metallic ground state was observed instead.\cite{Aerts} 
%Apart from The predicted semiconductors and corresponding energy gaps in Fig. \ref{alat1},
%agree with LDA+U results on the rare earth nitrides by Larson {\it et al.},\cite{Larson} with the exception of TmN and
%YbN that these authors predict to be semi-conductors. 
%For the remaining rare earth pnictides, only few unconfirmed semiconductors have been reported.
All the divalent monochalcogenides turn out to be semiconductors, in good agreement with
experiment.
The divalent character implies that in the chalcogenide semiconductors
the energy gap is brought about by localizing an
additional $f$-electron, as
depicted through the transition from (F) to (F') in Fig. \ref{schem},
in variance to the energy gap observed in the trivalent nitride semiconductors, resulting from the
filling of the unoccupied anion $p$-states, as represented
by scenario (A) in 
Fig. \ref{schem}. 
In the SIC-LSD the occupied 4$f$-levels tend to be situated at too low energies,\cite{Petit_AcOx} as this 
ground state methodology can not reproduce the spectroscopy of an eventual experimentally observed mid-gap $f$-level.
For EuO, EuS, and YbS, the calculations overestimate the energy gaps by a factor 2.\cite{Steeneken,Jayaraman_RS}

The dominant part of the electronic phase diagram in Fig. \ref{alat1} 
consists of metallic compounds. The indicated differences in conductivities can be straightforwardly
related to the trends in electronic structure observed when moving along the diagonals in Fig \ref{schem}. 
In the pnictides, the closing of the semiconducting gap due to the onset of pnictogen $p$ - rare earth $sd$ overlap
gives at first rise to a vanishingly small DOS, resulting in the semi-metals CeX to PrX and GdX to DyX (light blue areas).
As the overlap gradually increases (due to decreasing electronegativity and/or increasing lanthanide contraction), 
the behaviour becomes fully metallic (dark blue areas), and eventually
in the "late" rare earths, with more and more of the $f$-states getting occupied, 
heavy fermion-like behaviour sets in (yellow-red areas).
%earth pnictides, SmX, EuX, TmX, and YbX, marked in yellow-red in the phase diagram. 
%on the degree overlap between pnictogen $p$ and rare earth $sd$ as shown in Fig \ref{schem}.  
%As shown in Fig \ref{schem}, as we move along the diagonal the semiconducting gap closes with the onsetting overlap 
%Thus CeX and PrX, as well as most of the pnictides from GdX to DyX, 
%qualify as semi-metallic (light-blue) given the associated vanishinly small calculated DOSs.
%The corresponding electronic structure is settled at the transition between the scenarios (A) and (B) 
%of Fig. \ref{schem}, with the onset of pnictogen $p$ - rare earth $sd$ overlap and disappearance of the semiconducting gap.
%Said overlap increases as we move along the diagonal, and the behaviour becomes 
%fully  metallic (dark blue). As a result of the lanthanide contraction, more and more of the rare earth $f$-states
%start being occupied (scenario (C) of Fig. \ref{schem}) 
%eventually we observe  heavy fermion-like in most of the "late" rare 
%earth pnictides, SmX, EuX, TmX, and YbX, marked in yellow-red in the phase diagram. 
The rare earth chalcogenides display less variety in terms of metallic behaviour. 
Here we start out with a partially occupied conduction band, and semi-metallic behaviour 
is not observed. Instead, as can be seen from Fig. \ref{alat1}, the rare earth systems, CeX to PmX, and GdX to ErX (apart from the polonides and
a couple of tellurides) are
metallic.
Compared to the pnictides, in the late rare earth chalcogenides, the $f$-states start filling even before they have moved
far enough towards lower binding energies to achieve significant hybridization. Instead of forming a heavy metal,
a relatively pure $f$-state pins the Fermi level (scenarios (E) and (F) in Fig. \ref{schem}) and the divalent
semiconducting scenario (scenario (F') in Fig. \ref{schem}) correspondingly becomes energetically more favourable. 

The agreement between observed and predicted properties is overall very good, except maybe
for the calculated heavy fermion pnictides. 
%that emerges from the SIC-LSD calculations
%is in overall very good agreement with the experimental evidence.
%Regarding the pnictides, the "early" compounds, such as CeX and GdX, are indeed observed to be
%semi-metallic,\cite{Yamada} 
Thus the "late" rare earths, SmX, and YbX, are described as low carrier semi-metallic compounds,
seemingly at odds with the large DOS at the Fermi level
that is predicted in the SIC-LSD calculations. The experimentally observed large specific heat coefficient in these systems
has been interpreted as heavy fermion behaviour by some authors,\cite{Ott} whilst others
have explained it in terms of the Kondo effect.\cite{Monnier_YbX,Wachter_Kondo}
Since our calculations are based on the effective one electron
picture,
they can not adequately describe the many-body physics of these compounds. 
However the large peak at the Fermi level that is calculated in the
trivalent configuration indicates strong $pf$-mixing, i.e. a prerequiste for the hybridization between
the narrow $f$-peak and the sea of conduction electrons that underpins the Kondo effect. 
EuSb, YbBi, EuBi (as well as EuAs) do not actually crystallize in the NaCl structure investigated here,\cite{Horne} 
i.e., the very same pnictides
that we find to be divalent or close to divalent do not occur in nature.
As indicated schematically in Fig. \ref{schem}, for the pnictides, 
localization of an additional $f$-electron ((C) $\rightarrow$ (C') transition) implies the removal of
some bonding $p$-states. This, however, would destabilize the underlying NaCl crystal structure and, as a result of this
competition between band formation and correlations, it seems that these materials prefer to distort and
crystallize in a different structure altogether.

In Fig. \ref{alat1} the compounds marked as "intermediate valent" (IV) are those where the calculations find
close to degenerate energy configurations, meaning either $\Delta E_{II-III}$ or $\Delta E_{II-III}$ is smaller than 5 mRy.
Given the dual nature of the $f$-electrons, it is to be expected that
these materials are highly sensitive to environmental changes such as pressure\cite{Svane_SmX_EuX_statsolidi,Svane_PressureRE} or doping.
In the phase diagram of Fig. \ref{Etot} they correspond to the trivalent-divalent degenerate compounds in and around the 
white area, and to CeN, the only compound that is at the borderline of the tetravalent-trivalent degeneracy.
For the Tm chalcogenides, it emerges from XPS measurements that
TmS is metallic trivalent, TmSe is intermediate valent, and TmTe is semiconducting divalent.\cite{Bucher,Kinoshita} This, as can be seen
from Fig. \ref{alat1}, exactly
matches the valency sequence that we have derived from the SIC-LSD calculations.
For SmS we find $\Delta E_{II-III}$ = -5.6 eV, indicating a divalent semiconductor, but given the small energy difference, susceptible to
undergo a localization-delocalization transition under pressure,\cite{Svane_SmX_PRB} as has been observed experimentally.\cite{Henry}
%From the calculations
%we find SmS to be divalent, with $\Delta E_{II-III}$ = -5.6 eV, and increased pressure will lead to increased
%$f$-electron delocalization, pushing the compound towards
%intermediate valence, as the trivalent configuration gradually becomes competitive.
Apart from EuO, the rare earth monoxides do not appear to occur naturally, but a number of them have been synthesized under high pressure,
where it has emerged that EuO and YbO are divalent semiconductors, CeO, PrO and NdO are trivalent metals,
whilst SmO is a metal with the Sm in an intermediate valent state.\cite{Leger_RO} These findings, including the intermediate valent nature of SmO,
are again in excellent agreement
with our predictions.

In summary, this effective one-electron SIC-LSD methodology accurately predicts the phase transitions associated 
with the $f$-electron localization/delocalization. However, we should note here that because of the %neglected 
many-body effects, beyond those included in the SIC exchange-correlation energy functional, the representation 
of the calculated densities of states in terms of the described above phase diagram is most likely not the full 
picture of the underlying electronic structure. This is in particlular true in the 'IV' and heavy fermion areas
of the phase diagram, where those 'left out' many-body interactions will introduce additional refinements into 
the physical trends based on SIC-LSD. Therefore, compounds that are situated in those areas provide the greater 
challenge and should be further investigated by methodologies such as the dynamical mean field theory 
(DMFT)\cite{Pourovskii} or GW.\cite{Chantis_RX}

\section{Conclusions and outlook}

%It should come as not that large a surprise that the SIC-LSDA is doing so well for the rare earth compounds. 
%It transpires that a combined method  of Borje Johansson agrees very well with the present SIC-LSD \cite{Temmerman_YbX}. 
%The combined method takes the RE correlation energy from experimental atomic data and treats the condensed matter part 
%of the trivalent to divalent energy difference from ab initio calculations. The excellent agreement between the two 
%methods in the calculation of the energy differences between trivalent and divalent configurations shows that the 
%SIC-LSD total energy functional describes this energy difference sufficiently accurate. 
%
We have shown that the orbital dependent DFT approach, based on the LSD energy functional, appended by two important
corrections, namely the self-interaction correction and the tetrad effect, can provide a global picture
of the electronic structure of the rare earth monopnictides and monochalcogenides, expressed through the uncovered
trends in their behaviour. We believe that this SIC-LSD based study provides
an important understanding of the physical properties of these compounds. Although for a given compound the DMFT+LDA or GW approaches 
can give added insight into the details of the electronic structure, they can not at the present time be employed
for large scale predictive studies given that they either depend on parameters 
and/or are very time consuming.

Our study has given rise to a unified picture of the ground state valency and valency transitions across
the entire range of rare earth monopnictides and monochalcogenides, starting from tetravalency, through
trivalency to divalency, as one moves from the early to late rare earths, in correlation
with the ligands. We have also discovered the important links between the underlying electronic structure, 
valency and physical properties of these compounds, ranging from semiconducting to semi-metallic, metallic
and even heavy-fermion-like behaviour.

The present study has also established that if the energy differences between the energetically relevant valency
scenarios are small, then fluctuations will play an important role, giving rise to intermediate ground states. 
This effect can be described within the SIC-LSD formalism by mapping the valence (and also spin) fluctuations on
to disorder in the spirit of the Hubbard III approach,\cite{Lueders_LSIC} which we intend to apply in future
to the "intermediate" compounds identified in the present study.

\section*{Acknowledgments}
This work made use of computational resources at the Danish Center for Scientific Computing (DCSC) and NW-GRID
computers, the latter jointly provided by Daresbury Laboratory and the Universities of Lancaster, Liverpool
and Manchester with funding from the North West Development Agency.
We gratefully acknowledge helpful discussions with M. S. S. Brooks.

%\bibliography{leon_ref}
%\bibliography{/Users/leonpetit/desktop/lp/ARTICLES/BIBTE/leon_ref}

\end{document}